\documentclass[journal=jacsat,manuscript=article]{achemso}

\usepackage{chngcntr}
\usepackage{etoolbox}

\newcounter{mainfigure}
\newcommand{\beginsupplement}{
    \setcounter{mainfigure}{\value{figure}} 
    \renewcommand{\thefigure}{S\arabic{figure}} 
    \setcounter{figure}{0} 
}
\usepackage{siunitx}
\usepackage[version=3]{mhchem} 
\usepackage{bm}
\usepackage{fixltx2e}
\usepackage{amsmath}
\usepackage[colorinlistoftodos]{todonotes}
\definecolor{color1}{RGB}{127, 201, 127}  
\definecolor{color2}{RGB}{253, 191, 133}  
\definecolor{color3}{RGB}{56, 108, 176}   
\definecolor{color4}{RGB}{191, 91, 23}    
\definecolor{color5}{RGB}{102, 102, 102}  

\author{Mateusz Zelent}
\email{mateusz.zelent@amu.edu.pl}
\affiliation[UAM]
{Faculty of Physics and Astronomy, Adam Mickiewicz University in Poznan, Uniwersytetu Poznańskiego 2, PL-61-614 Poznan, Poland}

\author{Maciej Krawczyk}
\affiliation[UAM]
{Faculty of Physics and Astronomy, Adam Mickiewicz University in Poznan, Uniwersytetu Poznańskiego 2, PL-61-614 Poznan, Poland}

\author{Konstantin Y. Guslienko}
\affiliation[UPV]
{Depto. Polimeros y Materiales Avanzados: Fisica, Quimica y Tecnologia, Universidad del País Vasco, UPV/EHU, 20018 San Sebastian, Spain}
\affiliation[EHU] {EHU Quantum Center, University of the Basque Country, UPV/EHU, 48940 Leioa, Spain}
\affiliation[IKERBASQUE]
{IKERBASQUE, the Basque Foundation for Science, 48013 Bilbao, Spain}

\title[Multistable Skyrmions]
{
Beyond fixed-size skyrmions in nanodots: switchable multistability with ferromagnetic ring
}
\abbreviations{DMI, PMA, Sk, Co/Pd}

\begin{document}

\begin{abstract}
We demonstrate a novel approach to control and stabilize magnetic skyrmions in ultrathin multilayer nanostructures through spatially engineered magnetostatic fields generated by ferromagnetic nanorings. Using analytical modeling and micromagnetic simulations, we show that the stray fields from a Co/Pd ferromagnetic ring with out-of-plane magnetic anisotropy significantly enhance N\'eel-type skyrmion stability in an Ir/Co/Pt nanodot, even without Dzyaloshinskii-Moriya interaction. Most notably, we observe a multistability phenomenon, where skyrmions can be stabilized at two or more distinct equilibrium diameters depending on the ring's magnetization orientation. These stable states exhibit energy barriers substantially exceeding thermal fluctuations at room temperature, suggesting practical applications for robust multibit memory storage. By tuning geometric parameters of the ferromagnetic ring, we demonstrate precise control over skyrmion size and stability, opening pathways for advanced spintronic nanodevices.
\end{abstract}


\clearpage

\section{\label{sec:introduction}Introduction}

Magnetic skyrmions, nanoscale swirling spin textures\cite{nagaosa2013topological, Finocchio2021MagneticApplications}, can be stabilized at room temperature in chiral magnets and magnetic multilayer films exhibiting strong interfacial Dzyaloshinskii–Moriya interactions (DMI) \cite{fert2017magnetic}. These particle-like spin configurations display ultra-low critical currents for motion, rendering them promising information carriers for high-density devices such as the proposed racetrack memories and logic gates, wherein data are encoded by the presence or absence of individual skyrmions \cite{sampaio2013nucleation, fert2017magnetic}. The combination of nanometer-scale dimensions, topological protection, and efficient electrical manipulability positions skyrmions as attractive building blocks for next-generation spintronic memory and computing technologies \cite{everschor2018perspective, Finocchio2021MagneticApplications}.

However, a significant challenge for practical skyrmion-based devices lies in ensuring their robust stability under ambient, field-free conditions. In many known skyrmion-hosting materials, skyrmions are stable only within a narrow range of low temperatures or require continuous external magnetic fields~\cite{Moreau-Luchaire2015SkyrmionsMultilayers, Jiang2017SkyrmionsMultilayers}. Isolated skyrmions in single-layer films are often metastable, prone to collapse or elongate into stripe domains in the absence of a stabilizing field~\cite{Grebenchuk2024TopologicalFerromagnet}. This reliance on external magnetic fields complicates device integration and increases power requirements. Furthermore, conventional skyrmions typically possess a single equilibrium size determined by material parameters~\cite{Castro_Skyrmion_core_size_2016, Wang2018ASize}, offering only a binary state (presence or absence of a skyrmion) for information storage~\cite{Luo2021SkyrmionApplications}. While useful, this binary nature limits the stored information density and potential functionality of skyrmion-based devices.

Achieving multiple stable skyrmion configurations (i.e., multistability) within the same nanostructure could enable multi-level memory cells or novel logic states~\cite{Rzeszut2019Multi-bitJunctions,Hafliger2003ACell}. Such precise control over the skyrmion states has yet remained elusive. Although skyrmions with higher-order topological winding numbers (e.g., "target" or multi-turn skyrmions) were observed in specific bulk chiral magnets, suggesting the possibility of multi-state topological textures, switching between these states is non-trivial. Consequently, stable multi-state skyrmions were not realized in practical device geometries. This gap underscores the need for innovative methods to enhance the skyrmion stability and unlock additional stable states for advanced applications.

Recent studies \cite{Guslienko2018NeelNanodots,Tejo2017, Aranda2018MagneticInteraction} have provided crucial insights into magnetic skyrmions in confined geometries, highlighting the critical role of boundary conditions on skyrmion behavior and stability within ultrathin and multilayer nanodots. Our previous research \cite{PSSR:PSSR201700259} demonstrated the feasibility of bistable skyrmion states in multilayer nanodots with perpendicular magnetic anisotropy, where two distinct skyrmion sizes can be stabilized depending on initial conditions. Theoretical work \cite{Wang2018ASize} provides a quantitative understanding of how the skyrmion size and stability depend sensitively on a balance of the exchange, anisotropy, DMI, and Zeeman energies, confirming the energetic difficulty in stabilizing compact skyrmion states. Furthermore, Büttner et al. \cite{Buttner2018TheoryApplications} drew a crucial distinction between skyrmions primarily stabilized by DMI versus those stabilized by magnetic stray fields. Their analysis revealed that while DMI can, in principle, stabilize sub-10 nm skyrmions, achieving this at room temperature and zero magnetic field is extremely difficult in commonly used ferromagnetic multilayers (like Co-based systems). This often requires alternative materials (e.g., ferrimagnets) or non-zero applied fields. These papers highlight an urgent need for novel skyrmion stabilization mechanisms beyond the intrinsic DMI or external magnetic field, especially for realizing dense, field-free memory devices.

Leveraging engineered magnetostatic fields from adjacent magnetic layers offers another route for controlling skyrmions. Verba et al. \cite{Verba2018OvercomingMatrix} showed that dipolar coupling with a hard magnetic layer patterned as an antidot could stabilize magnetic vortex states in soft nanodots located under the antidots, significantly extending their stability range. In the recent work \cite{Zelent2023StabilizationNanostructures}, we explored a hybrid system consisting of a skyrmion-hosting nanodot placed on an in-plane magnetized soft ferromagnetic stripe. We found that the mutual magnetostatic interaction leads to significant effects: the skyrmion induces a magnetic imprint on the stripe, and the stray field from this imprint, in turn, acts back on the skyrmion. This interaction breaks the skyrmion's circular symmetry, leading to an asymmetric (egg-shaped) deformation, enhances its stability (allowing stabilization at lower DMI values), increases its overall size compared to an isolated dot, and even introduces skyrmion bi-stability within a specific DMI range \cite{Zelent2023StabilizationNanostructures}. This work highlighted how the magnetic stray fields in hybrid structures can profoundly modify skyrmion properties and potentially mitigate the skyrmion Hall effect. 

Building upon these insights – particularly the potential for manipulating skyrmions via engineered magnetostatic interactions from adjacent layers \cite{Verba2018OvercomingMatrix, Zelent2023StabilizationNanostructures} – we now investigate geometry designed nanostructures specifically to enhance skyrmion stability and achieve multistability. While the previous approaches used antidots or in-plane stripes to maintain robust skyrmion stability, achieving multiple controllable stable states in a simple, integrable structure remains a challenge. Therefore, there is strong motivation to develop a novel approach that not only stabilizes skyrmions (potentially without DMI) but also explicitly engineers multiple stable skyrmion states using magnetostatic fields.

In this work, we propose a novel design of a skyrmion hosting device, where a ferromagnetic ring above a multilayer nanodot is designed to achieve multistable skyrmions with enhanced stabilities. 
The magnetostatic stray field generated by this ring provides a stabilizing field within the nanodot's interior, acting as an integrated bias field analogous to an external magnetic field. This approach is 
particularly effective for stabilizing very small skyrmions (diameters $<$ 50 nm), with the spatially engineered magnetostatic field from the ring significantly deepening the skyrmion energy minimum, providing a robust barrier against its collapse, even in zero external field. 
Remarkably, we find that the interplay between the ring's stray field and the skyrmion leads to multiple stable skyrmion states. Specifically, for appropriate ring dimensions and magnetization, the system can support two distinct stable skyrmion configurations within the dot, "small-radius" skyrmion and "expanded" skyrmion, both corresponding to local energy minima. To our knowledge, this represents the first demonstration of multiple stable skyrmion sizes coexisting in the same nanostructure under identical conditions. Furthermore, we propose and numerically demonstrate the switching between both skyrmion states by applying short (below 0.5 ns) magnetic field pulses. 

The implications of these findings are significant for skyrmion-based spintronics. Enhanced stability ensures reliable information retention against thermal fluctuations and perturbations, crucial for memory and logic applications. 
Our ring-stabilized skyrmion design thus provides a practical pathway to harness multistable topological states in a simple geometry. In the following sections, we present a detailed analysis of the ring’s magnetostatic effect on skyrmion stability, map the conditions for multistability, and discuss the potential utilization of these multistable skyrmions in future high-density memory and unconventional computing applications.

\begin{figure*}[!htp]
    \centering
    \includegraphics[scale=1,width=0.70\textwidth]{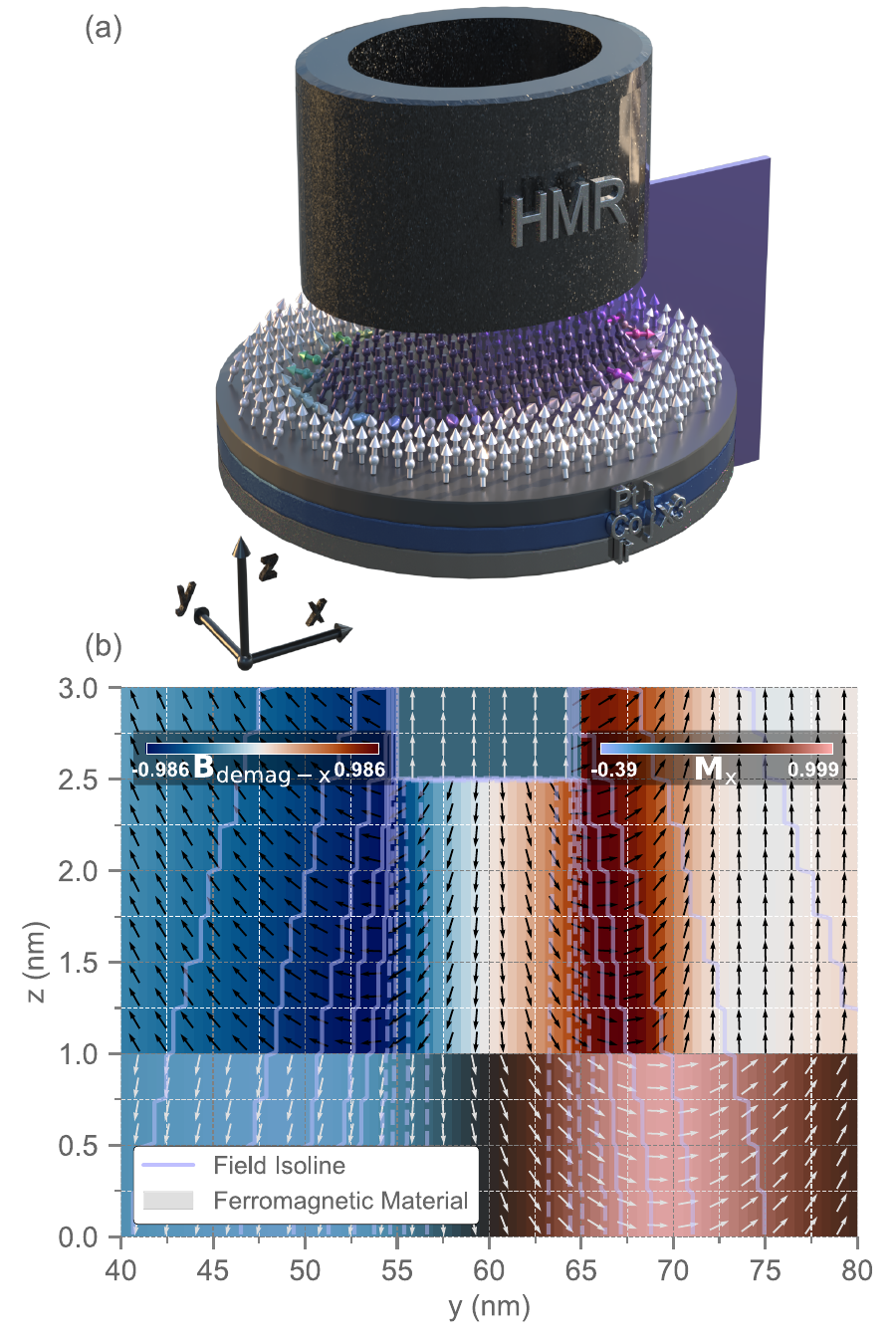}
    \caption{ 
    (a) Schematic representation of the proposed device consisting of Ir/Co/Pt multilayer nanodot  hosting the magnetic skyrmion, and Co/Pd ferromagnetic ring. Purple semi-transparent cuboid represents spatial cross-section of the magnetization texture and demagnetization field. (b) Composite visualization of the numerically calculated in-plane component of the demagnetizing field ($B_{\text{demag},x}$, left color map and black arrows) and the corresponding in-plane magnetization component ($M_x$, right colormap and white arrows), revealing steep field gradients at the inner and outer ring boundaries that induce pronounced magnetization tilting. Color scale represents field strength with red/blue indicating positive/negative values.
    }
    \label{fig:demag_field}
\end{figure*}

\section{Results and discussion}

Figure~\ref{fig:demag_field}(a) illustrates the geometry of our proposed device: Co/Pd multilayer ring (here the outer radius $r_2$: 65 nm, inner radius $r_1$: 55 nm, height: 10 nm) positioned 1.2 nm above (separated by nonmagnetic material) Ir/Co/Pt multilayer nanodot. 
By convention, we define the positive out-of-plane field as parallel to the skyrmion core magnetization direction. The nanodot has a fixed thickness $t$ of 1.2 nm with variable diameter, while the ring’s thickness and inner/outer radii serve as tunable parameters in our analysis. Detailed material parameters are provided in the Supplementary
Material (SM).

In investigations, we in-house developed version of Mumax3~\cite{A.TheMuMax3,Leliaert2014}, called AMUmax~\cite{MathieuMoalicAMUmax}, for the micromagnetic simulations, which solves the Landau-Lifshitz-Gilbert equation including magnetostatic, exchange, anisotropy, Zeeman, and DMI fields. To gain deep physical insight into the stability of N\'eel skyrmions within the ultrathin magnetic nanodot, complementing our micromagnetic simulations, we developed an analytical model based on a variational approach. Details of both methods are presented in SM.

The spatially non-uniform stray field generated by the ferromagnetic ring, whose magnetization is stabilised by the perpendicular magnetic anisotropy [see Fig.~\ref{fig:demag_field}(b), and Figs. S1 in SM], can create strategically positioned energy wells in the nanodot that can host and stabilize skyrmions. 

When this magnetostatic field aligns with the skyrmion's internal magnetization rotation (its chirality), it produces either a potential well (when parallel) or an energy barrier (when antiparallel) at specific locations relative to the ring. 
Performed analysis reveals that the skyrmion domain wall preferentially stabilizes near regions where the in-plane component of the stray field reaches its maximum.
This spatial correlation between field gradients and skyrmion domain wall positions is exploited in the following part of the paper to demonstrate enhanced stability and multistability of the skyrmion. Furthermore, as a distinct ring-like pattern of the stray field mirrors the geometry of the HMR, it offers a tool for designing a preferential skyrmion radius.

A key advantage of employing a ring configuration rather than a full disk lies in the asymmetric stray field distribution created by the distinct inner and outer edges of the ring. Each edge produces opposing effects on the energy landscape in the nanodot: one edge raises the energy while the other lowers it, creating sharply defined potential wells. This asymmetry significantly enhances skyrmion stability, particularly for small skyrmions that are typically prone to collapse. The steep energy barriers formed at specific radii prevent accidental relaxation to the single-domain state, effectively increasing the energy barrier against thermal fluctuations. Such precise control over stability regions would be unattainable with a uniformly magnetized disk, where field gradients would be present only at the outer radius.

\subsection{DMI-free skyrmion stabilization}

\begin{figure}[!htp]
\centering
\includegraphics[scale=1,width=\columnwidth]{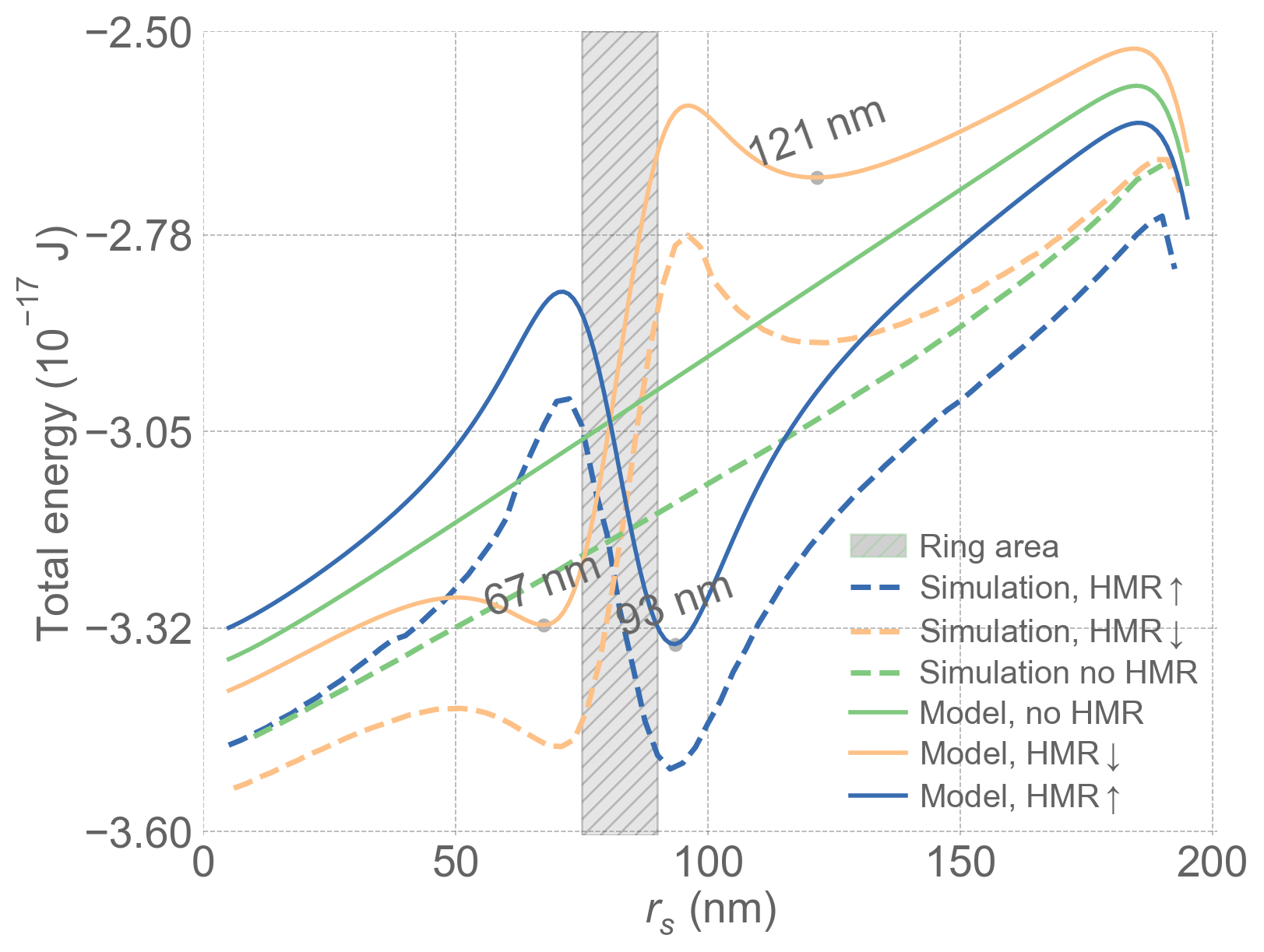}
3 \caption{Total magnetic energy of the N\'eel skyrmion as a function of its radius \( r_s \). The shaded region indicates the spatial extent of the HMR, with inner and outer radii of \( r_1 = 75.0 \) nm and \( r_2 = 90.0 \) nm, respectively. Solid lines represent the analytical model, while dashed lines correspond to micromagnetic simulations. The green curve denotes the reference case without the HMR for a nanodot of thickness \( t = 1.2 \) nm and zero DMI. The blue and orange curves correspond to cases where the HMR is magnetized perpendicularly downward (\(\downarrow\)) and upward (\(\uparrow\)), respectively.}
\label{fig:0dmi}
\end{figure}

We start the analysis from the reference system, i.e., the nanodot without the HMR (solid green curve in Fig.~\ref{fig:0dmi}) and zero DMI. Here, the nanodot exhibits no energy minima, thus no stable equilibrium for a skyrmion, confirming inherent skyrmion instability without DMI. Conversely, introducing the HMR ($r_1=75$ and $r_2=90$ nm) results in clear energy minima, effectively stabilizing skyrmions at specific diameters, in both magnetization orientations in the HMR (the blue and orange lines). Both analytical calculations (solid lines) and micromagnetic simulations (dashed lines) consistently confirm this stabilization mechanism, and confirm that our approach accurately captures the essential physics of magnetostatic skyrmion stabilization. The analytical model (the DeBonte skyrmion ansatz) simplifications under-represent slight edge effects in confined nanodots (Fig.~\ref{fig:skyrmion_profile}), resulting in minor discrepancies in the magnetic energy $\approx 0.2 \times 10^{-17}$ J (see supplementary Fig S1).
Thus, Fig.~\ref{fig:0dmi} demonstrates a remarkable finding: a N\'eel skyrmion is stabilized purely by the magnetostatic stray field from the HMR in the absence of DMI. 

The stabilization mechanism strongly depends on the direction of the HMR magnetization relative to the skyrmion core magnetization. For downward magnetization ($\downarrow$, blue curves), the HMR's stray field aligns with the skyrmion magnetization rotation across its entire radial profile (see Fig.1 (b) Fig.S2 (a-d) in SM). This complete compatibility creates a pronounced energy minimum, accompanied by a subsequent energy barrier due to opposite magnetization alignment at larger radii. The resulting deep energy well, corresponding to a skyrmion radius of approximately $r_s \approx 93$ nm, yields an exceptional stability barrier ($\Delta E = 0.4 \times 10^{-17}$ J) against thermal fluctuations at room temperature.

In contrast, upward magnetization ($\uparrow$, orange curves) yields two distinct stable skyrmion diameters at approximately 67 nm and 121 nm, i.e., inside and outside of the ring edges, respectively. In this scenario, the stray field from the HMR partially opposes the skyrmion core magnetization, creating a substantial energy barrier that prevents skyrmion contraction below a certain radius. Consequently, two clearly separated minima emerge: the inner minimum resulting from partial alignment with the skyrmion boundary, and the outer minimum due to partial alignment with the skyrmion core. Despite the fundamentally different stabilization processes, the depths of these minima—and thus their thermal stabilities—are comparable, underscoring the robustness of magnetostatic stabilization.

\subsection{Multistable skyrmion states}

\begin{figure*}[!htp]
\centering
\includegraphics[scale=1,width=\textwidth]{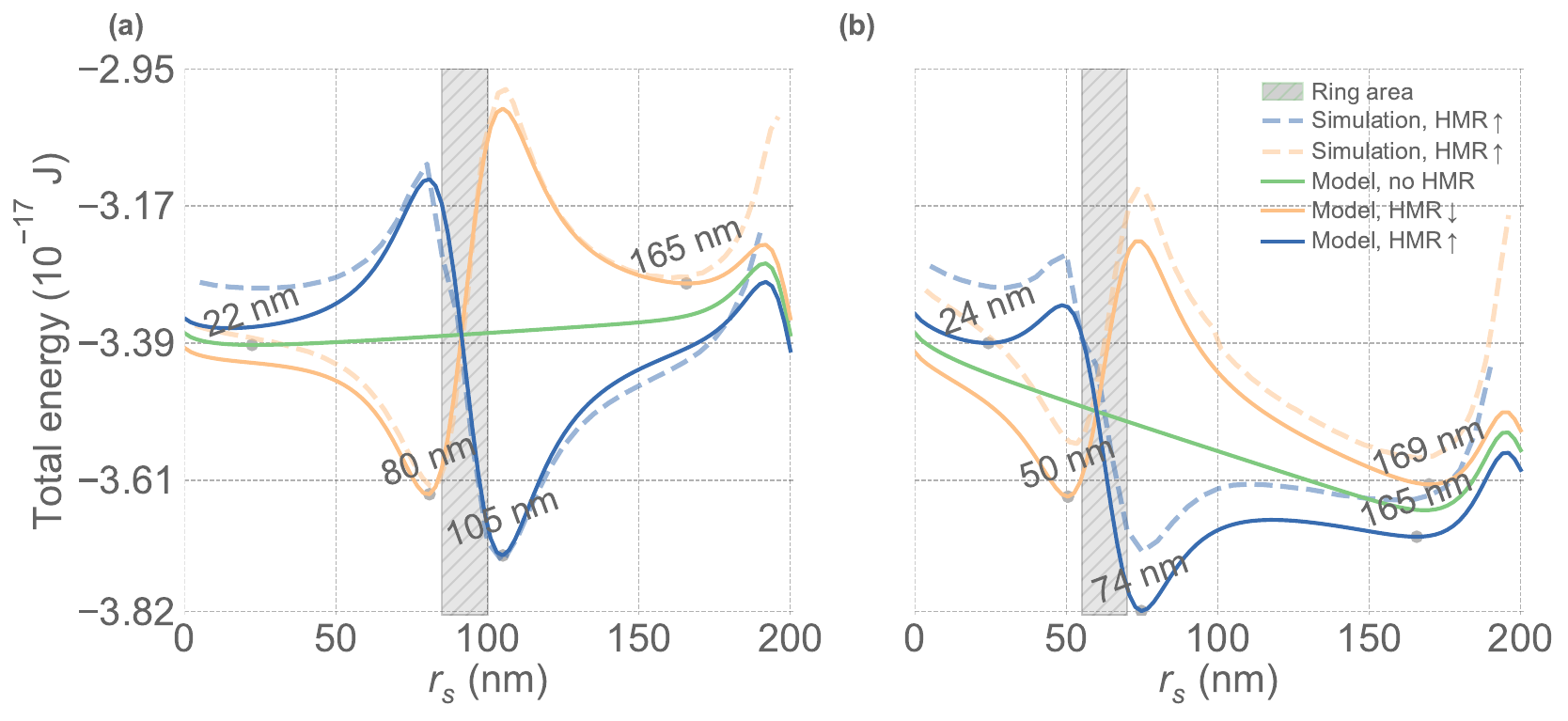}
\caption{
Total energy of the N\'eel skyrmion as a function of its radius \( r_s \), comparing analytical model predictions (solid lines) with micromagnetic simulations (dashed lines) for different HMR polarization and DMI strengths. Calculations were performed for a nanodot of thickness $t$ = 1.2 nm. Gray dots indicate energy minima, with corresponding approximate skyrmion radii labeled. The shaded gray region marks the radial extent of the HMR. The green curve denotes the reference case without the HMR. (a) DMI strength $D = 1.80$ mJ/m$^2$ with HMR dimensions \( r_1 = 85.0 \) nm, \( r_2 = 100.0 \) nm. With HMR magnetized upward (\(\uparrow\), blue lines), minima appear at \( r_s \approx 22 \) nm and \( r_s \approx 105 \) nm. With HMR magnetized downward (\(\downarrow\), orange lines), minima are observed at \( r_s \approx 80 \) nm and \( r_s \approx 165 \) nm.
(b) DMI strength $D = 2.60$ mJ/m$^2$ with HMR dimensions \( r_1 = 55.0 \) nm, \( r_2 = 70.0 \) nm. The reference case (green line) shows a preference for larger skyrmions due to stronger DMI. With HMR magnetized upward (\(\uparrow\), blue lines), minima appear at \( r_s \approx 24 \) nm and \( r_s \approx 165 \) nm. With HMR magnetized downward (\(\downarrow\), orange lines), a distinct minimum is observed at \( r_s \approx 50 \) nm, with another likely minimum at \( r_s \approx 169 \) nm (model) or larger radii.
}
\label{fig:multistable}
\end{figure*}

Having established that skyrmions can be stabilized in a nanodot solely through magnetostatic interactions with the HMR, we next examine the more general case where both DMI and the HMR stray field contribute simultaneously. This scenario has significant practical relevance for spintronic applications, as most heavy-metal/ferromagnet interfaces naturally exhibit some degree of DMI. 
Fig.~\ref{fig:multistable} presents the main effects of introducing DMI in combination with the HMR-induced magnetostatic field. The calculations were performed for a nanodot of thickness $t = 1.2$ nm with two different DMI strengths: $D = 1.8$ mJ/m$^2$ (panel a) and $D = 2.60$ mJ/m$^2$ (panel b). The shaded regions indicate the spatial extent of the HMR, with inner and outer radii varying between the two configurations: (a) $r_1=85$ nm, $r_2=100$ nm, and (b) $r_1=55$ nm, $r_2=70$ nm.

In Fig.~\ref{fig:multistable}(a), we observe distinct energy landscapes for different HMR polarization states. For downward ($\downarrow$) ring magnetization (orange curve), we find multiple energy minima at approximately 80 nm, and 165 nm. These minima are separated by energy barriers exceeding $\Delta E = 0.75 \times 10^{-17}$ J, which is sufficient to ensure thermal stability at room temperature (the thermal energy at 300 K is approximately $k_BT \approx 4.14 \times 10^{-21}$ J). Conversely, with upward ($\uparrow$) ring magnetization (blue curve), the energy landscape exhibits two stable positions at approximately 22 nm (shallow minimum), 105 nm (deep minima with $\Delta E$ exceeding $0.4 \times 10^{-17}$ J). Importantly, for the reference sample, i.e., a nanodot without HR (green line), we observe only one shallow minimum at 22 nm.

Fig.~\ref{fig:multistable}(b) demonstrates how increasing the DMI strength to $D = 2.60$ mJ/m$^2$ while adjusting the ring dimensions modifies the energy landscape. With downward HMR polarization, stable skyrmion states appear at 50 nm and 169 nm. For upward polarization, we observe three stable states at 24 nm, 75 nm, and 165 nm demonstrating clear multistability. The reference case without HMR (green curve) exhibits only a single minimum at 167 nm, highlighting how the ring-induced magnetostatic field fundamentally transforms skyrmion stability.

These results demonstrate that the interplay between DMI and the spatially varying magnetostatic field from the HMR creates complex energy landscapes, which can be designed to possess multiple well-defined minima. The positions and depths of these minima—and consequently the stable skyrmion diameters—can be precisely engineered by controlling the ring dimensions and its magnetization direction as well as DMI of the nanodot. This remarkable multistability emerges from the competition between the chirality imposed by the DMI and the effective chirality induced by the magnetostatic field from the HMR. When these chiralities align, existing energy minima deepen, and new metastable states may emerge. Conversely, when they oppose each other, certain stability points are suppressed while others are enhanced.

\subsection{Skyrmion state switching}
\begin{figure*}[!htp]
    \centering
    \includegraphics[scale=1,width=0.95\textwidth]{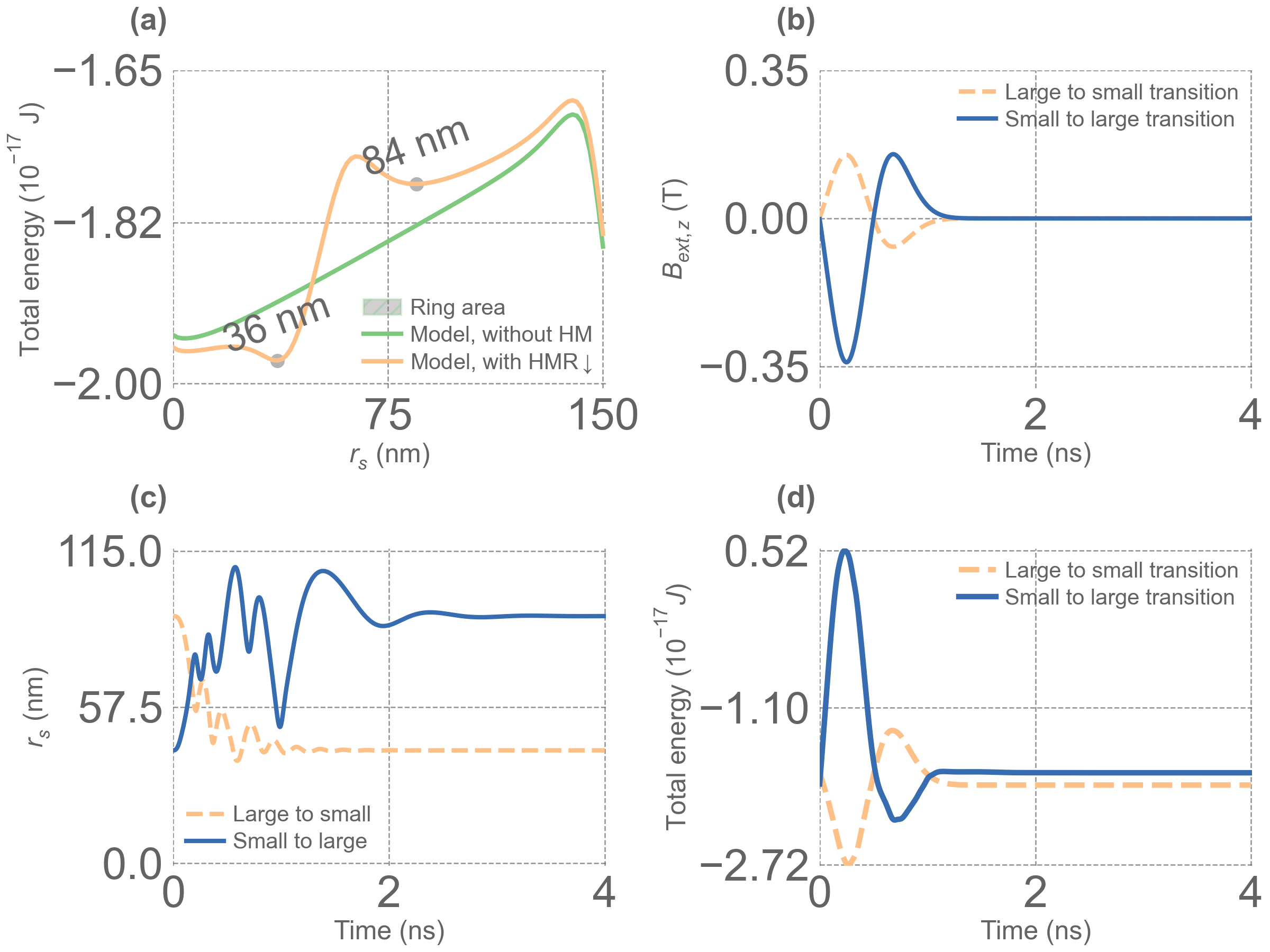}
    \caption{The switching between multistable skyrmion states using external magnetic field pulses. The simulations correspond to system parameters ($D=\SI{1.20}{mJ/m^2}$, $t=\SI{1.2}{nm}$, $\delta=1.20$, HMR $r_1=\SI{40.00}{nm}$, $r_2=\SI{60.00}{nm}$, $L_r=\SI{3.60}{nm}$) resulting in stable states at approximately $r_s = \SI{36.0}{nm}$ (small radius state) and $r_s = \SI{84.0}{nm}$ (large radius state).
    (a) Static energy landscape showing two distinct energy minima corresponding to the small and large radius skyrmion states, indicated by gray dots. The shaded region marks the radial extent of the HMR. The orange line represents the calculated energy profile using the model with the HMR.
    (b) Applied out-of-plane external magnetic field pulses ($B_{\text{ext},z}$) as a function of time. The solid blue line shows the negative pulse ($\approx \SI{-0.34}{T}$) used to trigger the 'Small to large transition', and the dashed orange line shows the positive pulse ($\approx \SI{+0.15}{T}$) used to trigger the 'Large to small transition'. Pulses have a duration of approximately $\SI{1.0}{ns}$.
    (c) Time evolution of the skyrmion radius ($r_s$) during the switching processes initiated by the pulses shown in (b). The solid blue line depicts the expansion from the small ($\approx \SI{36}{nm}$) to the large radius ($\approx \SI{84}{nm}$), while the dashed orange line shows the contraction from the large to the small radius. Relaxation to the stable state occurs within approximately $\SI{2.1}{ns}$.
    (d) Time evolution of the total system energy during the switching processes, corresponding to the radius dynamics shown in (c). 
    The energy rapidly changes during the pulse application and then relaxes towards the minimum energy value for the respective final state. All simulation were performed for $\alpha=0.1$.
    }
    \label{fig:switching}
\end{figure*}

Having established the existence of multistable skyrmion states in a nanodot, we now address the critical question of how to reliably switch between these states in a controlled manner—a prerequisite for practical memory and logic applications. Figure~\ref{fig:switching} illustrates our findings on skyrmion state switching mechanisms and dynamics.

For this demonstration we select the energy landscape shown in Fig.~\ref{fig:switching}(a), featuring multiple local minima (for the system characterized with following parameters: $D=\SI{1.20}{\text{mJ/m}^2}$, $t=\SI{1.5}{\text{nm}}$, $\delta=\SI{1.20}{\text{nm}}$, HMR $r_1=\SI{40}{\text{nm}}$, $r_2=\SI{60}{\text{nm}}$, $L_z=\SI{3.6}{\text{nm}}$), which creates a pathway for deterministic switching between the skyrmions of different diameters. The identified stable states correspond to skyrmion radii of $r_s \approx \SI{36.0}{\text{nm}}$ (small skyrmion, S) and $r_s \approx \SI{84.0}{\text{nm}}$ (large skyrmion, L), separated by an energy barrier ($\Delta E = 0.2 \times 10^{-17} \, \text{J}$). 

To investigate transitions between these states, we use short external magnetic field pulses applied out-of-plane ($B_{\text{ext},z}$) with varying amplitudes and durations.
Transition from the small-diameter state (S) to the large-diameter state (L) requires a negative field pulse that temporarily reduces the effective anisotropy, allowing the skyrmion to expand. We found that the pulse amplitude needed for this transition is approximately \SI{-0.15}{T} with a duration of about \SI{1.0}{ns}.
The optimized field pulse is shown in Fig.~\ref{fig:switching}(b) with the solid blue line. Conversely, switching from the L to S state requires a positive field pulse (dashed orange line) of approximately \SI{+0.35}{T} and the same duration (\SI{1.0}{ns}) that compresses the skyrmion.

It is noteworthy that the specific temporal profiles of the magnetic field pulses shown in Fig.~\ref{fig:switching}(b) are not simple square waves. This tailored shaping stems from the complexity of driving the skyrmion's size change across the energy barrier created by the HMR. The pulse profile typically features an initial phase with a strong rising or falling trend to initiate the transition, followed by a phase with an opposing trend (e.g., returning towards zero or slightly overshooting). This counter-phase is intentionally designed to mitigate the skyrmion's inertial response, effectively damping the resulting oscillations (like the breathing mode) and thereby shortening the relaxation time to the new equilibrium state. The necessary pulse duration and the subsequent relaxation time are inherently dependent on the Gilbert damping parameter ($\alpha$) present in the nanodot. The pulse profile itself can be optimized for a specific damping value to minimize switching time and energy. While the process warrants further optimization, our aim here was to demonstrate the feasibility of such controlled switching using shaped pulses.

The resulting switching dynamics are detailed in Fig.~\ref{fig:switching}(c) and (d). Figure~\ref{fig:switching}(c) shows the time evolution of the skyrmion radius ($r_s$) during the switching processes initiated by the pulses shown in Fig.~\ref{fig:switching}(b). It clearly depicts the radius changing from the initial value ($\approx \SI{36}{\nano\meter}$ or $\approx \SI{84}{\nano\meter}$) to the final value ($\approx \SI{84}{\nano\meter}$ or $\approx \SI{36}{\nano\meter}$, respectively). The process involves a rapid change of the skyrmion radius during the pulse duration, followed by damped oscillations around the new equilibrium state. Relaxation to the stable state occurs within approximately $\SI{2}{\nano\second}$ for the parameters and pulses used here. 

Figure~\ref{fig:switching}(d) presents the corresponding time evolution of the total system energy. The energy rapidly changes during the pulse (as the system is excited) and then relaxes towards the minimum energy value for the respective final state, also exhibiting oscillations consistent with the skyrmion radius dynamics.

The total switching time, including relaxation, is approximately $\SI{1.5}{\nano\second}$ to $\SI{2.0}{\nano\second}$. This compares favorably with conventional magnetic memory technologies, positioning skyrmion-based multistable memory as a competitive candidate for high-speed, high-density storage applications, especially considering the potential for further optimization of the switching protocol.

These findings collectively establish a comprehensive framework for controlling multistable skyrmion states in nanodot-ring hybrid structures, paving the way for advanced spintronic devices that exploit the unique properties of skyrmion multistability for multistate memory and logic applications.

\section{Conclusion}
Our study, which combines analytical modeling and micromagnetic simulations, shows that the stray field from a ferromagnetic ring with out-of-plane magnetic anisotropy can stabilize N\'eel skyrmions within an adjacent circular thin nanodot, even in the absence of Dzyaloshinskii-Moriya interaction. Furthermore, we show that the ring creates a complex energy landscape in which the stable skyrmion can exist with two or even more different diameters. The specific skyrmion stable states can be controlled by the magnetization polarity of the ring.
Moreover, we demonstrate that transitions between these distinct stable skyrmion states can be triggered and controlled using nanosecond external magnetic field pulses. The shape of this pulse can be designed to overcome the HMR-induced energy barriers and suppress the skyrmion anihilation, highlighting the feasibility of practical write operations in potential multistable memory or logic devices. 

Our finding expands the material landscape available for skyrmion devices, a versatile tool for tailoring skyrmion dimension, paving the way for multi-level data storage, enabling access to multistable states. This opens avenues for developing high-density, multi-level magnetic memory, reconfigurable logic elements, and potentially novel neuromorphic computing concepts.

\section{Acknowledgments}

The research was supported by the National Science Centre of Poland, project no. UMO-2023/49/B/ST3/02920.
K.G. acknowledges support by IKERBASQUE (the Basque Foundation for Science). The research of K.G. was funded in part by the Spanish Ministry of Science, Innovation and Universities grant PID2022-137567NB-C21 /AEI/10.13039/501100011033, by the Basque Country government under the scheme “Ayuda a Grupos Consolidados” (Ref. IT1670-22).



\appendix
\beginsupplement
\section{Supplementary Materials}
\subsection{\label{sec:analitical_model}Analytical model}

To gain physical insight into the stability of N\'eel skyrmions within the ultrathin magnetic nanodot, complementing  micromagnetic simulations, we develop an analytical model based on a variational approach. This model first establishes the energy landscape for an isolated skyrmion in the circular dot, providing a baseline upon which the effects of the external HMR stray field can be later assessed.

The system's geometry is illustrated in Fig. 1(a), consisting of an Ir/Co/Pt multilayer nanodot hosting the skyrmion, surrounded by a Co/Pd ferromagnetic ring with strong out-of-plane uniaxial anisotropy. By convention, we define positive out-of-plane field as parallel to the skyrmion core magnetization direction. The nanodot has a fixed thickness of 1.2 nm with variable diameter, while the ring's thickness and inner/outer radii serve as tunable parameters. Detailed material parameters are provided in the Methods section.

The total magnetic energy density of our system is given by:
\begin{equation}
e(\mathbf{m}) = A(\nabla \mathbf{m})^2 + e_{\text{DMI}}(\mathbf{m}) - K_u m_z^2 + e_m(\mathbf{m}),
\end{equation}
where $A$ is the exchange stiffness constant, $K_u$ is the uniaxial anisotropy constant, $e_{\text{DMI}}$ represents the interfacial DMI energy density, and $e_m$ corresponds to the magnetostatic energy density.

For a thin circular magnetic dot with radius $R$ and thickness $L_{d} = 1.2$ nm, we parameterize the magnetization using the unit vector $\mathbf{m} = \mathbf{m}(\Theta, \Phi)$, where the spherical angles $\Theta$ and $\Phi$ depend on the polar coordinate vector $\mathbf{\rho} = (\rho, \phi)$ in the dot plane. The total energy is given by $E[\mathbf{m}] = L_d \int d^2 \rho \, e(\mathbf{m})$.

The DMI energy density, takes the form:
\begin{equation}
e_{\text{DMI}} = D[\mathbf{m}_z(\nabla \cdot \mathbf{m}) - (\mathbf{m} \cdot \nabla)m_z]
\end{equation}
where $D$ is the interfacial DMI parameter. The \(m_z\) is the z-component of magnetization. 

The magnetostatic energy is generally non-local, but within the limit of an ultrathin dot (\(\beta = L_d/R \ll 1\)), it can be simplified and expressed in local form as:
\begin{equation}
e_m(\mathbf{m}) = \frac{\mu_0 M_s^2 m_z^2}{2}
\end{equation}

Therefore, we account for the effective anisotropy via a renormalized uniaxial anisotropy constant \( K = K_u - \mu_0 M_s^2 / 2 \). Additionally, we define the magnetic material quality factor as \( Q = 2K_u / \mu_0 M_s^2 \), and we consider systems where \( Q \geq 1 \).

For axially symmetric skyrmions, the magnetization angle $\Theta$ depends only on the radial coordinate $\rho$, $\Theta = \Theta(\rho)$ and $\Phi = \phi + \phi_0$, where the helicity $\phi_0 = 0$ or $\pi$ for N\'eel skyrmions. The skyrmion topological number is given by $N = [\cos(\Theta(0)) - \cos(\Theta(R))] / 2$. We assume $\tan[\Theta(r)/2] = \exp[-f(r)]$ and use the DeBonte ansatz~\cite{Debonte1973PropertiesPlatelets} \( f(r) = \ln\left(\frac{r}{r_s}\right) + \frac{1}{\delta}(r - r_s) \) to reproduce skyrmion profile, where we define the skyrmion radius $r_s=R_{s}/l$ as the distance where $m_z(R_s) = 0$ and $l=\sqrt{A/K}$, corresponding to $\Theta(R_s) = \pi/2$ or $f(r_s) = 0$. Parameter $\delta$ corresponds to the skyrmion wall width. The magnetization components within anzatz are \( m_{z}(r) = \cos\Theta(r) = \tanh(f(r)) \) and \( m_{\rho}(r)=\sin\Theta(r) = \frac{1}{\cosh(f(r))} \)

We calculate the skyrmion energy from Eq. (1-3) to determine regions of skyrmion stability and metastability. A skyrmion state is defined as stable when it has the lowest energy (ground state) compared to other magnetization states, and metastable when its energy is higher than that of another magnetization configuration but separated by an energy barrier. We get the expressions of energy versus skyrmion radius, which allows us to calculate the radially symmetric skyrmion energy as:
\begin{equation}
E(r_{s}) = 2\pi A L_d \int_0^{R_d} \rho \left[
\left( \frac{1}{\rho^2} + 1 \right) m_\rho^2 +
\left( \frac{1}{\delta} + \frac{1}{\rho} \right)^2 m_\rho^2 +
d \left( -\left( \frac{1}{\delta} + \frac{1}{\rho} \right) m_\rho + \frac{1}{\rho} m_\rho m_z \right)
\right] d\rho - \pi A L_d R  _d^2,
\end{equation}

where the radial coordinate $\rho$, dot radius $R_{d}$ and skyrmion wall width $\delta$ are in units of $l$. 

\begin{figure}[!h]
    \centering
    \includegraphics[scale=1,width=0.95\columnwidth]{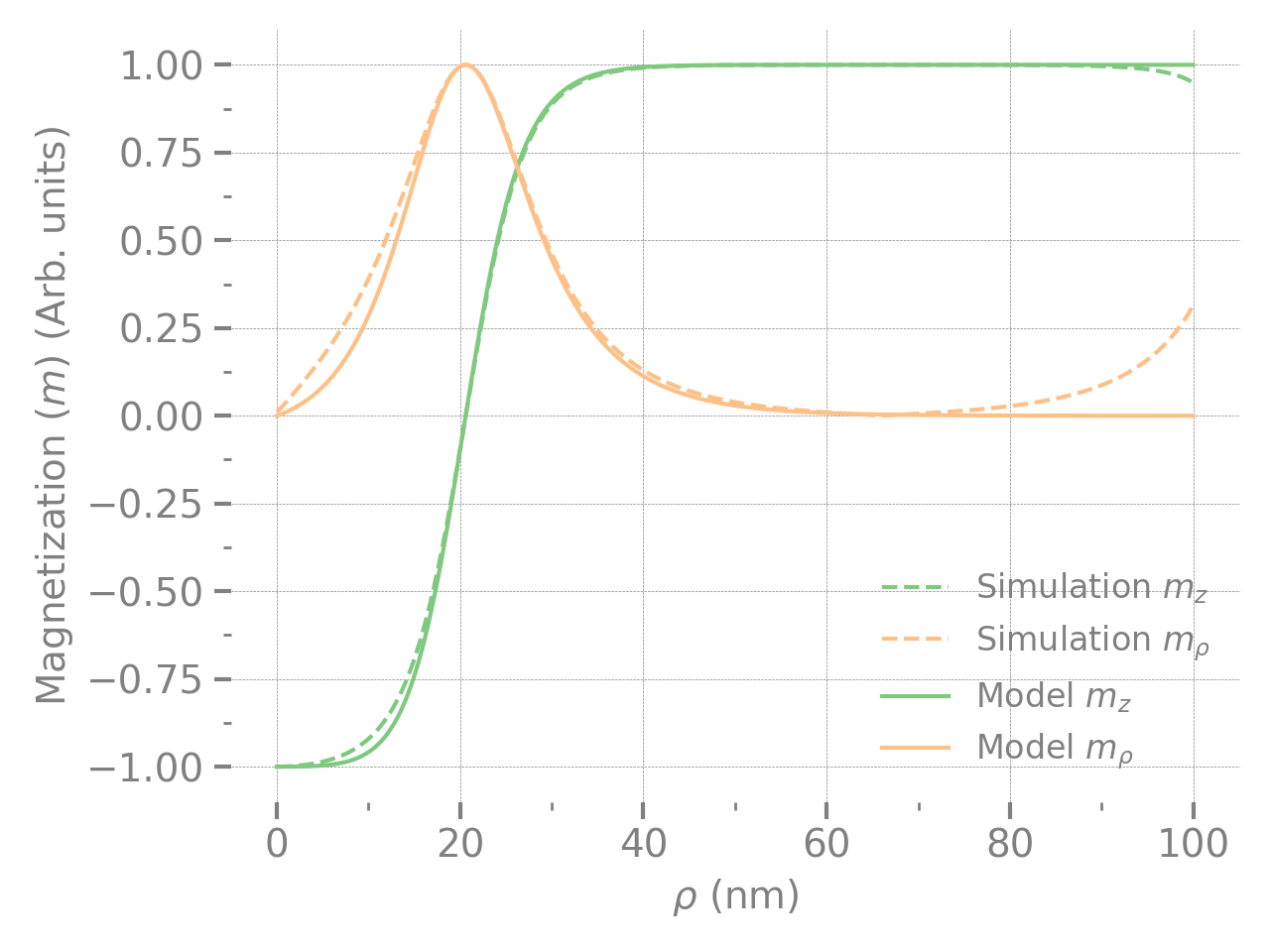}
    \caption{(a) Comparison of simulated and modeled 20 nm radius N\'eel type skyrmion magnetization profiles in an ultrathin multilayer nanodot. The graph shows the in-plane (orange lines) and out-of-plane (green lines) components of the magnetization as a function of the radial coordinate $\rho$. The dashed lines correspond to micromagnetic simulations, while the solid lines represent the analytical model based on the DeBonte ansatz.}
    \label{fig:skyrmion_profile}
\end{figure}

Figure \ref{fig:skyrmion_profile} compares this analytical approximation with micromagnetic simulations, demonstrating excellent agreement for the magnetization profile in the central region of the nanodot. The DeBonte ansatz accurately captures both the in-plane and out-of-plane magnetization components as functions of the radial coordinate, though it shows minor deviations near the edges because boundary effects (arising due to finite value of the DMI energy) not fully accounted for in the analytical model. This model has been previously applied to isolated nanodots as reported in Refs.~\cite{Aranda2018SingleFilms,Tejo2017}.


\subsection{Ring stray magnetic field}
The second crucial component of our model addresses the interaction between the skyrmion and the magnetostatic stray field $\mathbf{H}_{\text{ring}}$ generated by the ferromagnetic ring. This Co/Pd multilayer ring with strong perpendicular magnetic anisotropy is characterized by its outer radius $R_{\text{out}}$, inner radius $R_{\text{in}}$, and thickness $L_{\text{r}}$, serving as a local source of magnetic field that fundamentally modifies the skyrmion energy landscape.

The stray field generated by the HMR can be decomposed into two principal components: (a) the out-of-plane component \( H_z(\rho, z) \), which effectively modifies the perpendicular anisotropy in the nanodot, and
(b) the in-plane radial component \( H_{\rho}(\rho, z) \), which is non-uniform and reverses direction near the inner and outer edges of the ring.

The influence of these field components on skyrmion stability can be understood through their interaction with the skyrmion magnetization. The out-of-plane component \( H_z \) directly modifies the effective anisotropy, either enhancing or reducing it depending on the relative orientation of \( H_z \) and the skyrmion core magnetization. Meanwhile, the radial component \( H_{\rho} \) introduces a torque that effectively acts as an additional DMI-like contribution, influencing the skyrmion diameter and chirality. In particular, \( H_{\rho} \) reaches maxima near the inner and outer radii of the ring, with opposite signs, creating regions where the skyrmion wall energetically prefers to localize.

The total energy of the system, incorporating the interaction with the stray field, becomes:
\begin{equation}
\begin{aligned}
E_{\text{total}}(r_{s}) &= E(r_{s}) - \mu_0 M_{sd} \int_0^{R_d} \int_{0}^{L_d} 2\pi\rho \left[ m_{\rho}(\rho) H_{\rho}(\rho, z) + \right. \\
&\left. m_{z}(\rho) H_{z}(\rho, z) \right] dz d\rho,
\end{aligned}
\end{equation}
where the first term $E(r_{s})$ is given by Eq. (4), representing the intrinsic skyrmion energy in the absence of the ring, and the second term describes the Zeeman-like interaction energy between the skyrmion magnetization and the ring-generated magnetic field.

\begin{figure}[!htp]
    \centering
    \includegraphics[scale=1,width=0.9\columnwidth]{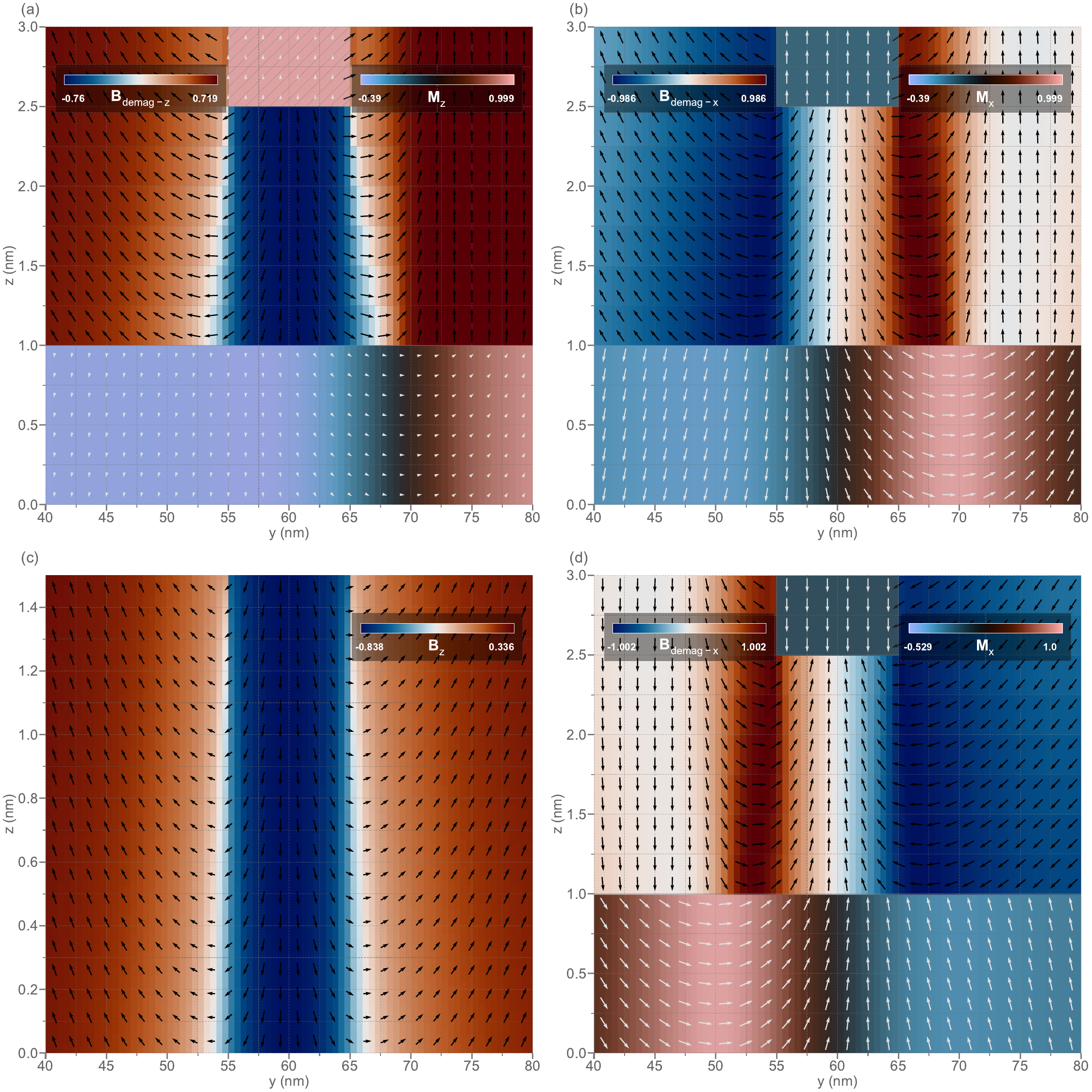}
\caption{
Spatial visualization of simulated magnetic fields (demagnetizing field $\mathbf{B}_{\text{demag}}$) and magnetization ($\mathbf{M}$) compared with an analytical calculation ($B_z$) in the y-z cross-section for different stable skyrmion states induced by the HMR. Colormaps indicate the magnitude of the specified component (see color bars). Black arrows represent field vector components ($B_y, B_z$ or $B_{\text{demag-y}}, B_{\text{demag-z}}$), while white arrows represent magnetization vector components ($M_y, M_z$) in the y-z plane.
(a, b) Simulated state for HMR magnetized upward ($\uparrow$), stabilizing a large-radius skyrmion. (a) Out-of-plane components: $B_{\text{demag-z}}$ (left, colormap with black field arrows) and $M_z$ (right, colormap with white magnetization arrows). (b) In-plane components: $B_{\text{demag-x}}$ (left, colormap with black field arrows) and $M_x$ (right, colormap with white magnetization arrows).
(c) Analytically calculated profile of the out-of-plane component ($B_z$) of the HMR stray field, with corresponding field vectors (black arrows). Note the different z-axis range compared to other panels.
(d) Simulated state for HMR magnetized downward ($\downarrow$), stabilizing a small-radius skyrmion, showing in-plane components: $B_{\text{demag-x}}$ (left, colormap with black field arrows) and $M_x$ (right, colormap with white magnetization arrows). Note the distinct visualization style for the region $z < 1.0$ nm in panels (b) and (d).
}
\label{fig:s2}
\end{figure}

To calculate the spatial distribution of the stray field components, we employ a semi-analytical approach based on the Fourier-Bessel expansion of the magnetostatic potential. For a uniformly magnetized ring with perpendicular magnetization \( M_{sr}\), the field components are given by:
\begin{multline}
H_{\rho}(\rho, z) = M_{sr} \int_{0}^{\infty} J_1(k \rho) \left[ J_1(kR_{\text{in}}) - J_1(kR_{\text{out}}) \right] e^{-k|z|} (1 - e^{-kL_{\text{r}}}) k \, dk,
\end{multline}
\begin{multline}
H_z(\rho, z) = M_{sr} \int_{0}^{\infty} J_0(k \rho) \left[ J_1(kR_{\text{in}}) - J_1(kR_{\text{out}}) \right] e^{-k|z|} (1 - e^{-kL_{\text{r}}}) k \, dk,
\end{multline}
where \( J_n(x) \) are Bessel functions of the first kind, \( k \) is the wavevector in the Fourier-Bessel space, and the expressions account for the finite thickness of the ring through the factor \( (1 - e^{-kL_{\text{r}}}) \). The calculated spatial cross-section below the ring is shown in Figure~S\ref{fig:s2} (a,b,d~-- simulations, c~-- analytical calculations). 

As demonstrated in Fig. 1(c-d), the resulting spatial distribution of these field components creates characteristic patterns that profoundly influence skyrmion stability. The $H_z$ component forms a circular band aligned with the ring position, while the $H_{\rho}$ component exhibits sharp transitions with sign changes at the inner and outer edges of the ring. These field gradients generate effective potential wells at specific radii within the nanodot, leading to the discrete stable skyrmion diameters observed in our calculations and simulations.

The competition between the ring-induced field and the intrinsic skyrmion energetics (exchange, DMI, anisotropy) determines the final stable or metastable states. By systematically varying the ring parameters—thickness $L_{\text{r}}$, inner radius $R_{\text{in}}$, and outer radius $R_{\text{out}}$—we can engineer energy landscapes with desired stability characteristics. Furthermore, by reorienting the ring magnetization, we can dynamically modify the energy landscape, enabling switching between different stable skyrmion diameters as demonstrated in the main text.

\subsection{\label{sec:numerical_approach}Numerical simulations}
To investigate the relationship between skyrmion diameter and its stability in an ultrathin circular magnetic dot under varying out-of-plane external magnetic fields, we employ a micromagnetic approach. The physical system under study is illustrated in Fig. 1 (a), where we define our convention: a positive out-of-plane magnetic field aligns parallel to the skyrmion core, while a negative field is applied in the opposite direction. The dot features a variable diameter $d$ with a fixed thickness of 1.2 nm. We consider IrCoPt materials with the parameters taken from the literature~\cite{PSSR:PSSR201700259,Moreau-Luchaire2016AdditiveTemperature}. We use the following material parameters for a nanodot: saturation magnetization $M_\mathrm{s}$ = 956 kA/m, exchange stiffness constant $A_{\text{ex}} = 10$ pJ/m, $D$ $-$ changes in a range from 0 to 2 mJ/m$^2$, out-of-plane magnetic anisotropy constant $K_{\mathrm{u}} = 0.8$ MJ/m$^3$ and Gilbert damping constant $\mathrm{\alpha} =$ 0.1. For HMR, we assume Co/Pd with the following magnetic parameters: $M_\mathrm{s} = 810$ kA/m, $A_{\mathrm{ex}} = 13$~pJ/m, out-of-plane magnetic anisotropy constant $K_{\mathrm{u}} = 0.45$ MJ/m$^3$  and $\mathrm{\alpha} = 0.1$. The studied system was discretization uniformly with 0.75 $\times$ 0.75 $\times$ 1.2 nm$^{3}$ unit cells to precisely imitate the rounded geometries with high accuracy. 

The micromagnetic simulations are performed by using the Mumax3\cite{Leliaert2018FastMumax3} which solves the Landau-Lifshitz-Gilbert equation:
\begin{align}
 \frac{\text{d}\mathbf{m}}{\mathrm{d}t}=\gamma \mu_0 \frac{1}{1+\alpha^{2}} (\mathbf{m} \times \mathbf{H}_{\mathrm{eff}}) + \alpha \mu_0 \left( \mathbf{m} \times (\mathbf{m} \times \mathbf{H}_{\mathrm{eff}}) \right),
 \label{eq:LLG}
\end{align}
where $\textbf{m} = \textbf{M} / \mathit{M}_{\mathrm{s}}$ is the normalized magnetization, $\textbf{\text{H}}_{\mathrm{eff}}$ is the effective magnetic field acting on the magnetization, $\gamma=-1.7595 \cdot 10^{11}$ Hz/T is the gyromagnetic ration, $\alpha$ is the Gilbert damping. 
In this paper, the following components were considered for the effective field $\textbf{H}_{\mathrm{eff}}$: demagnetizing field $\textbf{\text{H}}_{\mathrm{d}}$, exchange field $\textbf{\text{H}}_{\mathrm{ex}}$,
Dzyaloshinskii-Moriya exchange field $\textbf{\text{H}}_{D}$, and uniaxial anisotropy field $\textbf{\text{H}}_{\mathrm{Ku}}$. External magnetic field and thermal effects were neglected. Thus, the effective field $\textbf{H}_{\mathrm{eff}}$ is expressed as:
\begin{align}
 \textbf{H}_{\mathrm{eff}} =
 \textbf{H}_{\mathrm{d}} + \textbf{H}_{\mathrm{ex}} + \textbf{H}_{\mathrm{D}} + \textbf{H}_{\mathrm{Ku}},
\end{align}
where
\begin{align}
 \textbf{H}_{\mathrm{ex}} = 2 \frac{A_{\mathrm{ex}}}{\mu_0 \mathit{M}_{\mathrm{s}}} \Delta \textbf{m},
\end{align}
where $A_{\mathrm{ex}}$ is the exchange constant, and 
\begin{align}
\textbf{H}_{\mathrm{D}} = \frac{2 \mathit{D}}{\mu_0\mathit{M}_{\mathrm{s}}} 
\left(\frac{\partial m_z}{\partial x},\frac{\partial m_z}{\partial y},-\frac{\partial m_x}{\partial x},-\frac{\partial m_y}{\partial y}, \right).
\end{align}
 The uniaxial anisotropy is accounted in the form:
\begin{align}
\textbf{H}_{\mathrm{Ku}} =
\frac{2\textit{K}_{\mathrm{u}}}{\mu_0 \mathit{M}_{\mathrm{s}}} 
\left( \textbf{u} \cdot \textbf{m} \right) \textbf{u},
\end{align}
where $\textit{K}_{\mathrm{u}}$ is the first order uniaxial magnetic anisotropy constant and $\textbf{u}$ is a unit vector indicating the anisotropy direction. 

The multi-stable skyrmion states were computed by performing simulations of skyrmion stabilization as a function of $r_{s}$, assuming the two skyrmion relaxations, one when the initial state is a large-diameter skyrmion (75\% of the disk diameter) and the second with a small-diameter skyrmion (10\% of the disk diameter). 

To compute the energy landscape $E(r_s)$ as a function of skyrmion radius $r_s$ (e.g., dashed lines in Fig.~\ref{fig:0dmi}), we employed a \textit{frozen spins} technique adapted for micromagnetic simulations~\cite{PSSR:PSSR201700259}. For each target radius $r_s$, we define a narrow ring of width 1 cell size (0.75~nm) centered at the skyrmion edge. Within this ring, the magnetization is initialized such that $m_z = 0$ and the in-plane component points radially outward, i.e., $m_r = 1$, corresponding to a vortex-like configuration at the domain wall. The magnetization inside the ring is oriented antiparallel to that outside the ring, facilitating the formation of a Néel-type skyrmion with radius $r_s$. Only the spins within this narrow ring are constrained during the simulation; all other spins are free to relax. This enables the system to reach a local energy minimum for a fixed skyrmion radius without imposing any assumptions on the detailed skyrmion profile. Unlike the semi-analytical model, which requires a predefined ansatz, this method preserves full flexibility of the internal structure. Repeating the minimization over a range of radii $r_s$ yields the full energy profile $E(r_s)$, revealing stable and metastable configurations as well as energy barriers between them.

The stray field analysis (Fig.~\ref{fig:demag_field}b-d) involved calculating the field generated by the HMR in its uniformly magnetized state. The switching simulations (Fig.~\ref{fig:switching}) involved applying time-dependent external field pulses ($\mathbf{H}_{\mathrm{ext}}(t)$) and solving the full LLG equation (\ref{eq:LLG}), without frozen-spins technique, to observe the dynamic evolution of the skyrmion radius and system energy.

\clearpage
\bibliography{references,local}
\end{document}